# Supervised Link Prediction in Co-Authorship Networks Based on Author Node-Based Features

Doaa Hassan[1, 2, *], Mohammad Al Hasan[1, *]

1 Luddy School of Informatics, Computing and Engineering, Indiana University-Indianapolis, United States
2 Computers and Systems Department, National Telecommunication Institute, Cairo, Egypt
* Corresponding authors contacts:
doaa@nti.sci.eg
alhasan@iu.edu

**Abstract:**
Predicting the emergence of future research collaborations between authors in academic social networks (SNs) is a very effective example that demonstrates the link prediction problem. This problem refers to predicting the potential existence or absence of a link between a pair of nodes (authors) on the co-authorship network. Various similarity and aggregation metrics were proposed in the literature for predicting the potential link between two authors on such networks. However, the relevant research did not investigate the impact of similarity of research interests of two authors or the similarity of their affiliations on the performance of predicting the potential link between them. Additionally, the impact of the research performance indices of academic SN co-authors on link prediction performance was not highlighted. To this end, in this paper we propose an integrative supervised learning framework for predicting potential collaboration in co-authorship network based on similarity of the research interests and the similarity of the affiliations of each pair of authors in this network. Moreover, our proposed framework integrates the aggregation of research performance indices of each author pair and the similarity between the two authors nodes with the research interest and affiliation similarity as four features/metrics for predicting the potential link between each two authors. Our experimental results obtained from applying our proposed link prediction approach to the two largest connected graphs of two huge academic co-authorship networks, namely ArnetMiner and DBLP, show the great performance of this approach in predicting potential links between two authors on large-scale academic SNs.

## 1. Introduction

Social networks have an important role in communication among people from different groups or communities. Social network analysis (SNA) focuses on investigating the relation among members of social network and measuring the interaction and flow between those members [1]. The basic idea of SNA is to model the social network using graph theory as a graph (either directed or not directed) that has a set of vertices/nodes and edges/links among them. One of the great challenges in SNA is the link prediction problem. Link prediction aims to determine the potential interactions (i.e., potential links) among nodes of social networks which are possible to occur in the future. Link prediction has many applications [1, 2] such as helping to understand the mechanisms that trigger the evolution in a social network, friend recommendation, e-commerce, and bioinformatics. Our presented study in this paper focuses on solving the link prediction problem in academic co-authorship networks with a great performance for predicting the possible research collaboration that might occur among authors of those networks. We chose this type of SNs as a case study since the prediction of the establishing links/collaboration between authors of this network would help them to collaborate together in order to achieve their research goals [3].
The literature shows that supervised and unsupervised learning has been used for link prediction in academic SNs [4, 5]. In this paper, we focus on the link prediction using supervised learning techniques in which a supervised learning strategy [6] is employed for treating link prediction task as a binary classification problem (two author nodes establish a link in the form of co-authorship relationship or not). A pair of author nodes is defined as an instance of the classification problem in which a set of features are extracted to describe the author pair, while the target variable represents either a *positive label* that refers to the existence of a research collaboration (link) between the two author nodes in the future or a *negative label* that refers to the absence of such link or collaboration. Most of the supervised link prediction presented in the literature used features that are extracted by computing some metrics from exploring the structural patterns of the network [7]. Such metrics can be either local metric computed by examining the relationship between each node and its neighbors (such as common neighbors and adamic adar) [8, 9] or global metric that is computed over the whole entire network (such as Kats) [10].
We highlight here that there were also some existing studies that used other types of features for link prediction. An example of those features is the feature extracted by computing a sum function to aggregate the common keywords that represent the research interests of a pair of authors [4] or by computing the similarity between the context of two published papers by a pair of authors or their abstracts or their titles [11, 12]. In this study we extend this idea using the similarity between the research interests and affiliations of two authors as two features for link prediction. The former is determined by computing the similarity between the keywords that generally express their research interests regardless of the contents of the papers/abstracts that they wrote, while the latter is determined by measuring the text similarity between the affiliation texts of two authors. Additionally, we use other two types of features for link prediction including the sum of research performance indices of two authors and the similarity between the two nodes that represent both authors on co-authorship network graph. We call all four types of features author node-based features. The author node-based features proposed in this study for predicting potential collaboration between two authors in academic SN using supervised link prediction technique were introduced before in [13]. However, this approach was only tested on a very small subset of ArnetMiner [14] [15] co-authorship network with not necessarily connected vertices and it achieved a modest/poor performance of supervised link prediction. Therefore, our main contribution in the presented study is to show that the author node-based features presented in this paper lead to a very high performance of the supervised ML model for predicting potential links between authors of the co-authorship network when it is applied to the largest connected component graph of a huge academic SNs. The experimental results of this study on the largest connected component (LCC) graphs of two graphs representing large and well-known co-authorship networks, namely Arnetminer [14] [15] and DBLP [16] show the effectiveness of our approach for the link prediction using author node-based features in predicting the potential collaboration between authors of these networks. For each collaboration network, the author's information including his/her research interests, affiliation and author's research performance indices are extracted from AMiner-Author [17] that is publicly available for research purpose and saves information about authors including his/her affiliation, research interests, and research performance indices.

The rest of this paper is organized as follows: Section 2 discusses the related work. Section 3 presents our proposed supervised link prediction approach using author node-based features. Section 4 presents the performance evaluation results. Section 5 provides a discussion. Finally in Section 6 we conclude the paper and provide some directions for the future work.

## 2. Related work

Most of the research work introduced to solve the link prediction problems in academic SNs relies on the idea of extracting features from the collaboration network structure [18, 19]. However, our focus in this section is on the link prediction techniques that were proposed for predicting potential collaboration between two author nodes using features extracted from authors' nodes. Those features refer to the features that could describe the authors' nodes on the collaboration network such as their affiliation, research performance, papers that they wrote or their abstracts and keywords, etc in addition to the features generated from measuring the similarity between each two author nodes on the graph representing the SNs. Therefore, in the following we briefly state some of those techniques.
M. Hasan et el. [4] introduced a supervised link prediction approach that used a set of features extracted per author node pair from the topological structure of the network and keyword match count that expresses the common keywords that both nodes/authors used in their papers to solve the link prediction problem. In addition, they used some other aggregated features that were extracted for each author node pair such as sum keywords that represent research interests and sum of papers. They showed that a small subset of those extracted features had a very important role in solving the link prediction problem with a high performance.
T. Wohlfarth and R. Ichise [20] presented a new semantic non-structural feature called "Keywords match count" (KMC) that was extracted by comparing the main topic of research of two author nodes in order to decide about linking them. This was achieved by counting the number of words in common between all the titles of the previous papers of both authors to extract the new semantic feature. The titles were preprocessed for removing the stop words (e.g., "the, "or, "on, etc). Then the Jaccards coefficient was used to compare the similarity of two sets of words. In order to achieve more accuracy with their approach, they combined the KMC semantic feature with a feature that counts the number of events in common between researchers (represented by a journal where they both wrote a paper or a conference where they both presented their work).

Sachan and R. Ichise [21] presented a supervised learning method for building link predictors from semantics attributes of the nodes like title and abstract information. They introduced a new feature/variable called Abstract Keywords Match Count (AKMC). This feature is computed using the Jaccard's coefficient to calculate the similarity of two abstracts penned by a pair of authors. Next this feature is combined with other features extracted from the network structure and used to train a supervised link predictor.

Cohen et al. [22] proposed a keywords-based collaborator recommendation model for returning a ranked list of possible collaborators based on a set of given keywords. The collaborator recommendation problem was formalized by a given a query q that consists of a researcher s (a member of SN) and a set of keywords k (e.g., an article name or topic of future work). To solve this problem, the collaborator recommendation model should return a high-quality ranked list of possible collaborators for s on the topic k.

Liang et al. [23] proposed a method for cross-disciplinary collaboration recommendation by designing a cross-disciplinary discovery algorithm based on topic-modeling to extract potential research fields. They analyzed the research field correlations to discover the collaboration patterns with researchers' profiles. Next, they developed a recommendation algorithm to provide a specific recommendation list of potential research fields according to the discovered cross-disciplinary collaboration patterns.

P. M. Chuan. Son et el. [24] presented a supervised link prediction approach for predicting links between authors in co-authorship network based on measuring the similarity of two paper contents that were written by a pair of authors in that network. They proposed a new content similarity measure called LDAcosin that uses the Latent Dirichlet Allocation (LDA) method for extracting the content similarity between two papers, where the higher similarity indicates the higher possibility of link in the future between the two authors of the two papers. This new content similarity measure is used as a new metric feature for performing link prediction in the co-authorship network.

Makarov et al. [25] studied the problem of predicting collaborations in co-authorship network by formulating it as link prediction task on weighted co-authorship network. The authors on this network play the role of nodes, and weighted edges connecting two authors are formed by storing either a number or quality metric of research papers co-authored by these authors. The link prediction task was performed using regression machine learning model that was trained based on network features constructed using network embedding.

D. Hassan [13] introduced a supervised link prediction model for predicting potential links between authors of the academic co-authorship network. This model was trained with a combination of author content-based features and author node similarity features. The former includes the features extracted from computing the similarity between research interests of two authors, the similarity of their affiliation and the sum of their research performance indices. The latter includes the features extracted by computing the similarity between two author nodes on co-authorship network graph using common similarity measures such as dice, Jaccard and Inv. log weighted. However, the performance of this model was poor and was not tested on large academic collaboration networks.

Pradhan and Pal [26] introduced a multi-level fusion-based model for collaborator recommendation. They computed the author–author cosine similarity from abstracts and titles and then using that value to weigh edges in the author–author graph.

Our proposed approach extends the research directions above by introducing a supervised link prediction model that used author-content based features and author node similarity features for predicting potential collaboration between authors of the collaboration network. Both type of features were used for link prediction in [13]. However, our presented study uses both type of features in a different way from what was introduced in [13]. This is achieved by extracting both types of features from the two LCC graphs of the two graphs representing two large collaboration networks ArnetMiner and DPLB. The idea of using similarity between research interests and affiliation as a feature for predicting the potential collaboration between two author nodes is inspired by the idea of computing the author–author cosine similarity from abstracts and titles to weigh edges in the author–author graph proposed in [26]. However, our proposed approach applies the cosine similarity to the research interests and affiliations of two authors nodes instead of finding the similarity of the papers abstract and titles. Additionally, the idea of using the aggregation of research performance indices as a feature for link prediction was inspired by the aggregated features approach for link prediction introduced in [4] that were extracted for each author node pair such as sum of keywords that represent research interests and sum of papers. Finally, the similarity between two author nodes was commonly used in the literature as a feature generated from calculating the similarity score between two vertices in SN [27].

### 3. Proposed link prediction approach

Our proposed link prediction framework in this study (Figure 1) extracts author node-based features from the LCC graphs of the two graphs representing ArnetMiner or DPLB collaboration networks. Such implementation is considered as a refinement and improvement over our early implementation in [13] that only considered a non-LCC graph of a very small subset of the ArnetMiner collaboration network with only 50000 author nodes. Therefore, in this proposed study, we have applied various ML classifiers to the two LCC graphs of the two graphs representing ArnetMiner and DPLB collaboration network to predict the possible collaboration between two author nodes in each co-authorship network using the extracted author node-based features. Those features include the features extracted by computing the similarity between the research interests of each two author in the collaboration network (RI-sim), the similarity between their affiliations (AF-sim), the sum of their research performance indices (I-sum), and the similarity between the two nodes representing both authors on co-authorship network graph (Node-sim). These author node-based features (RI-sim, AF-sim, I-sum and Node-sim) are used to train the supervised ML classifiers used for collaboration prediction in the testing phase.

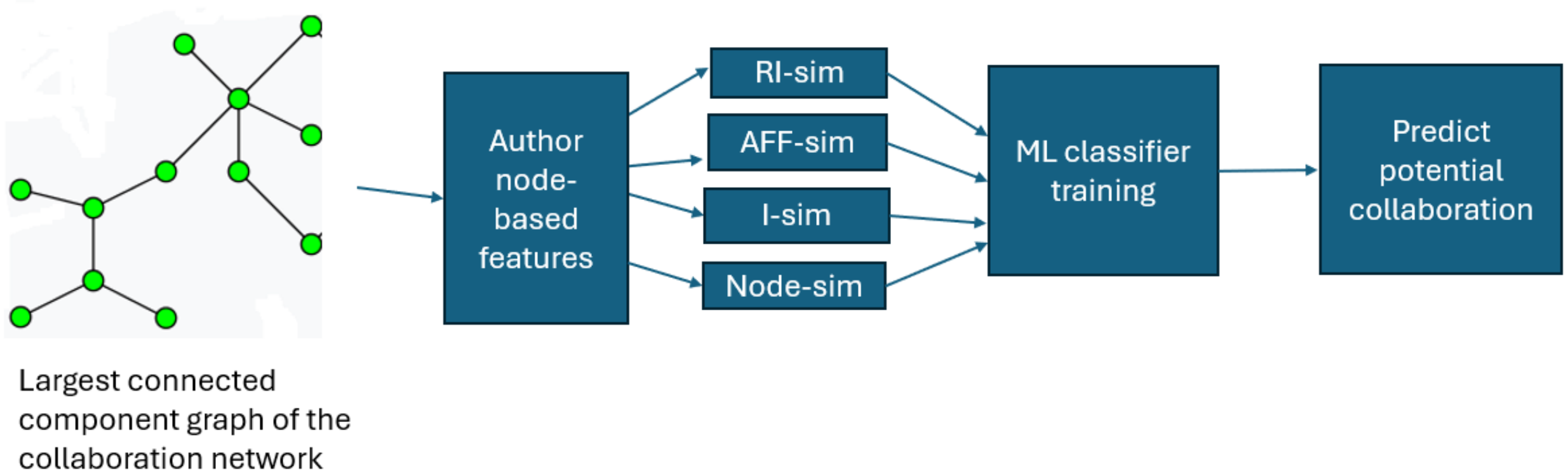

**Figure 1. The proposed supervised link prediction approach based on author content-based**

**3.1 Co-authorship network dataset**

The dynamic co-authorship networks used in this study are constructed from ArnetMiner dynamic co-authership datasets [14] and the co-authorship network of DPLB [16]. ArnetMiner dataset has 1,768,776 co-authorships that takes the form of publications published between 1986 and 2012 by 1,629,217 authors. DBLP co-authorship dataset is a computer science bibliography that was originally created at the University of Trier in 1993 to provides open bibliographic information on major computer science journals and proceedings, and it has 1,482,029 unique authors and 10,615,809 timestamped co-authorship edges between authors [28]. In the experimental approach of this study, we consider every author in the two LCC graphs of the two graphs representing ArnetMiner and DPLB co-authorship networks. For each author on any of the two networks, we extract his /her research interest from the AMiner-Author dataset [17] that saves information about authors including his/her affiliation, research interests, and research performance indices.

**3.2 Features based on similarity of research interests and affiliation**

The similarity of research interests would help in establishing links between two authors in collaboration network. This is because when two authors share research interests, they are more likely to have a research collaboration. Same affiliation or working places also can help authors to meet easily and increase the probability of starting up a new research collaboration. Our approach uses the *cosine similarity* measure known in the literature [29] for computing the similarity between research interests of two authors (RI-sim) and the similarity between their affiliations (AF-sim). The cosine similarity between two string texts $S_1$ and $S_2$ that represent the research interests or affiliations of two authors on the collaboration network is formally defined as follows:

$$Cosine(S1, S2) = \frac{S1.S2}{\parallel S1 \parallel \parallel S2 \parallel}$$

where each of $S1$ and $S2$ are first represented as two vectors and the cosine similarity measure is defined as the angle between the two vectors.

### 3.3 Features based on aggregating research performance indices

Each author in academic co-authorship network has a set of indices that evaluates his/her research performance (e.g., H-index and I index). Combining the research performance indices of two authors induces a new pattern (I-sum) for describing the author pair in academic co-authorship network. Such a pattern is used as a feature in our approach for predicting potential collaboration. Thus, our approach defines various aggregation functions corresponding to combination of diverse types of research performance indices for two authors nodes on the collaboration network. Each aggregation function aggregates one type of research performance indices for each pair of authors nodes on the collaboration network. For example, the aggregation function H-sum aggregates the h research performance indices of two author nodes. We have used the following research performance indices to evaluate the research performance of a researcher on the collaboration network.

- *pc-index* that represents the count of published papers by a researcher.
- *cn-index* that represents the total number of citations of a researcher.
- *h-index* that measures both the productivity and citation impact of researcher's publications.
- *pi-index* refers to P-index with equal A-index of an author, where P-index represents the productivity indexes (journal impact factors in which the researcher published his/her publication) and A-index computes the weighted total of peer-reviewed publications that refers to the collaboration performance (C- index) and P- index [30] .
- *upi-index* refers to the P-index with unequal A-index of a researcher.

### 3.4 Features based on node similarity

Those features are extracted from measuring the similarity score between two vertices on the graph representing the academic SN. Thus, in our proposed approach we extract those types of features by computing the similarity scores between each two nodes on the LCC graph of the graph representing either ArnetMiner or DPLB collaboration networks. In the following, we summarize the various metrics that we use for measuring the similarity score between two nodes on the graph:

**Jaccard coefficient:** The Jaccard coefficient is a similarity index computed between two author nodes in the collaboration network by counting the number of the common neighbors of the two nodes and divide it by the total neighbors of both nodes [31]. Formally the Jaccard coefficient between two author nodes u and v with a set of neighbors N(u) and N(V) respectively is defined as follows:

$$Jaccard(u,v) = \frac{N(u) \cap N(v)}{N(u) \cup N(v)}$$

**Dice similarity coefficient:** The dice is a similarity index computed between two author nodes in the collaboration network by counting the number of the common neighbors of the two nodes and divide it by the sum of their degrees d then multiply the result by two [32][1]. Formally the dice similarity score between two author nodes u and v with a set of neighbors N(u) and N(V) respectively is defined as follows:

$$dice(u,v) = 2\frac{N(u) \cap N(v)}{d(u) \cup d(v)}$$

***Inv.log-weighted* coefficient*:*** it is also known as Adamic/Adar similarity. It is a similarity index computed between two author nodes by calculating the sum of the inverse logs of the degrees d (x) of the common neighbors ($N(u) \cap N(v)$) of the two nodes where x ∈ ($N(u) \cap N(v)$) [33, 34]. Formally the *Inv.log-weighted* coefficient between two author nodes u and v with a set of neighbors *N(u)* and *N(v)* respectively is defined as follows:

$$Inv.log - weighted(u,v) = \sum_{x \in (N(u) \cap N(v))} \frac{1}{\log d(x)}$$

[1] The node degree is the number of connections that the node has to other nodes in the network.

### 3.5 Developing ML models for potential collaboration prediction

Seven supervised ML models from different categories have been developed using scikit-learn, a free ML python library [35] for predicting the potential links among authors in the dynamic co-authorship network. This includes Random Forest (RF), Decision Tree (DT), k-Nearest Neighbors (KNN), the Neural Network (NN), Naive Bayes (NB), Support Vector Machine (SVM) and Logistic regression (LR) [36]. We apply those models to the two LCC graphs of the two graphs representing ArnetMiner and DBLP co-authorship networks datasets. The sample in each dataset takes the form of an edge that is initially labeled based on the existing links between each two author nodes. Thus, the existing links in the co-authorship network refer to positive links with class value of 1, while the unlinked author node pairs in the network refer to negative links with class value of 0. Since there are many negative links in comparison to the positive ones, we randomly down-sample the number of negative links in each dataset to make it equal to the number of positive samples to obtain a balanced dataset of edge samples.

We choose the co-authorship that occurred between the year 1990 and the year 1993 in ArnetMiner dataset as the training dataset, while we choose the co-authorship that occurred between the year 1994 and the year 1995 as the testing dataset. For DBLP dataset, we choose the co-authorship that occurred between the year 1992 and the year 2010 as the training dataset, while we choose the co-authorship that occurred between the year 2011 and the year 2015 as the testing dataset. A visualization of training and testing datasets (in the form of LCC graphs) of ArnetMiner and DBLP co-authorship networks are shown in Figure 2 (a) and (b) respectively. Our approach solves the link prediction problem by predicting the new links appearing in ArnetMiner and DBLP testing datasets using the author node-based features extracted from the ArnetMiner and DBLP training datasets respectively.

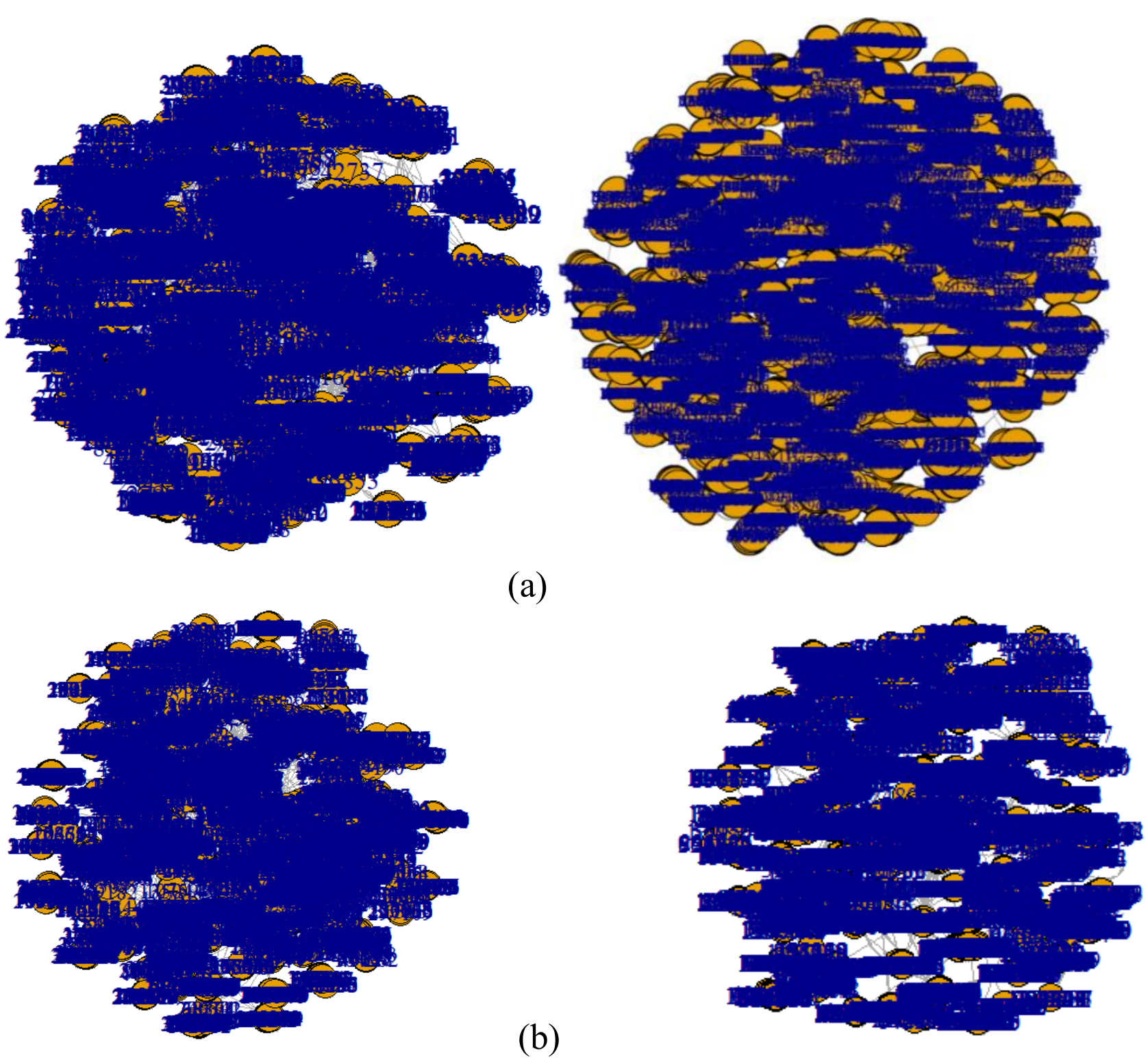

Figure 2. A visualization of experimental co-authorship networks (a) A visualization of training datasets for ArnetMiner and DBLP respectively (b) A visualization of testing datasets for ArnetMiner and DBLP respectively.

### 3.6 Performance evaluation metrics

We have reported about the performance of the developed supervised ML models for link prediction using different performance evaluation metrics including the accuracy, precision (P), recall (R), and F1-score (F) [36]. We also used the entire two-dimensional area under receiver operating curve (AUC) [37] which measures how accurately the model can distinguish between two things (e.g. determine if an edge is positive or negative). Below we introduce the mathematical equations for the first four metrics:

$$Accuracy = \frac{TP + TN}{TP + TN + FP + TN}$$

$$Precision = \frac{TP}{TP + FP}$$

$$Recall \; = \frac{TP}{TP + FN}$$

$$F1 - score = 2\frac{Precision * Recall}{Precision + Recall}$$

Where:

TP stands for true positive, and it represents the number of positive links that were correctly predicted as positive links.
TN stands for true negative, and it represents the number of negative links that were correctly predicted as negative links.
FP stands for false positive, and it represents the number of negative links that were misclassified as positive links.
FN stands for false negative, and it represents the number of positive links that were misclassified as negative links.

**4. Performance evaluation results**

Table 1 and 2 show the performance of various ML models when predicting potential links between author pairs on ArnetMiner and DBLP collaboration networks respectively using extracted author node-based features. All ML models were used with the default settings. Tables 1 shows that NN, LR, and NB achieve the best accuracy, precision, recall, and F1-score with a very slight outperformance for LR in terms of AUC on ArnetMiner dataset. Table 2 shows that NN, LR, NB, and RF achieve the best accuracy, precision, recall, and F1-score with a very slight outperformance for LR in terms of AUC on DBLP dataset. When comparing the results in Table 1 to Table 2, we notice that the performance of all ML link predictors on DBLP co-authorship network outperforms the performance of link prediction on ArnetMiner co-authorship network. This is due to the huge number of edges in DBLP in comparison to lower number of edges in ArnetMiner. In general, all ML link predictors achieve high performance of link prediction on either ArnetMiner or DPLP, which indicates that the author node-based features have good discriminating ability and a great impact on predicting the potential authors' collaborations.

**Table 1.** Performance of different ML models applied to ArnetMiner experimental dataset for link prediction using author node-based features.

| **Classifier** | **Accuracy %** | **Precision** | **Recall** | **F1-score** | **AUC** |
|---|---|---|---|---|---|
| **NN** | 84 | 1.00 | 0.67 | 0.80 | 0.835 |
| **LR** | **84** | **1.00** | **0.68** | **0.81** | **0.838** |
| **NB** | 84 | 0.99 | 0.68 | 0.81 | 0.837 |
| **KNN** | 79 | 0.90 | 0.65 | 0.76 | 0.792 |
| **SVM** | 76 | 0.67 | 1.00 | 0.80 | 0.755 |
| **RF** | 82 | 0.93 | 0.69 | 0.80 | 0.823 |

**Table 2.** Performance of different ML models applied to DBLP experimental dataset for link prediction using author node-based features.

| Classifier | Accuracy% | Precision | Recall | F1-score | AUC |
|---|---|---|---|---|---|
| **NN** | 99 | 0.99 | 0.99 | 0.99 | 0.989 |
| **LR** | **99** | **0.99** | **0.99** | **0.99** | **0.991** |
| **NB** | 99 | 0.99 | 0.99 | 0.99 | 0.989 |
| **KNN** | 97 | 0.98 | 0.95 | 0.97 | 0.966 |
| **SVM** | 98 | 1.00 | 0.96 | 0.98 | 0.979 |
| **RF** | 99 | 0.99 | 0.99 | 0.99 | 0.989 |

### 4.1 Author node-based features importance

We have reported about the importance of Author node-based features in terms of link prediction accuracy of the best performance classifiers (NN, LR, and NB) tested on ArnetMiner dataset and best classifiers (NN,LR, NB, and RF) tested on DBLP dataset with each single type of author node-based features. Table 3 and 4 show the importance of each single type of author node-based feature in terms of accuracy of the link prediction on ArnetMiner and DBLP respectively achieved using best performance classifiers when it is trained with each single type of feature.

**Table 3.** The performance of best ML models on link prediction in terms of accuracy (%) with a single type of feature when applied to ArnetMiner experimental dataset.

| Classifier | RI-sim | AFF-sim | I-sum | Node-sim |
|---|---|---|---|---|
| NN | 50 | 50 | 50 | 84 |
| LR | 50 | 50 | 51 | 84 |
| NB | 50 | 50 | 50 | 84 |

**Table 4.** The performance of best ML models on link prediction in terms of accuracy (%) with a single type of feature when applied to DBLP experimental dataset.

| Classifier | RI-sim | AFF-sim | I-sum | Node-sim |
|---|---|---|---|---|
| NN | 50 | 50 | 50 | 99 |
| LR | 50 | 50 | 50 | 99 |
| NB | 50 | 50 | 50 | 99 |
| RF | 50 | 50 | 50 | 99 |

As Tables 3, 4 show, the similarity between two authors node (Node-sim) is the best type of feature for predicting the link between two authors on either ArnetMiner or DBLP collaboration networks, while the similarity between the research interest and affiliation of two authors node (RI-sim, and Aff-sim) and the aggregation of the research performance indices of two authors (I-sum) have less impact on predicting the link between two authors on either ArnetMiner or DBLP collaboration networks with almost an equal feature importance.

### 4.2 Comparison with the performance of early link prediction implementation

Our link prediction approach introduced in this study is considered as a refinement and improvement for our early implementation of this approach introduced in [13]. Therefore, to show the outperformance of our approach in this study over the performance of our early implementation of this approach presented in [13], we compare the performance of best classifiers in our early implementation (RF and KNN) with performance of those classifiers in the current study when they are applied to ArnetMiner dataset. Table 5 shows this comparison. As the table shows, our current approach outperforms our early implementation in terms of precision, recall, F1-score and AUC, while it was not reported explicitly about the accuracy of link prediction in [13]. We do not provide the same comparison for DBLP dataset as our early implementation of link prediction was not applied and tested on this co-authorship network.

**Table 5.** Comparing the performance of link prediction in the current study to what already was introduced.

| Classifier | Approach | Accuracy% | Precision | Recall | F1-score | AUC |
|---|---|---|---|---|---|---|
| RF | Early in [13] | - | 0.202 | 0.319 | 0.233 | 0.682 |
| | Presented approach | 82 | 0.93 | 0.69 | 0.80 | 0.823 |
| KNN | Early in [13] | - | 0.148 | 0.287 | 0.213 | 0.648 |
| | Presented approach | 79 | 0.90 | 0.65 | 0.76 | 0.792 |

Table 6 provides the same comparison using each single type of feature. As the table shows, RF and KNN achieve better performance in terms of precision, recall, F1-score and AUC using our current approach than our early implementation of this approach when any classifier is trained with either RI-sim or Aff-sim or Node-sim as a type of features. However, the performance of RF and KNN achieved a slightly better performance using the sum of research performance indices of two potential collaborative authors (I-sum) in the early implementation in comparison to the performance in the presented study using this feature. This is because the early implementation reports I-sum as the best feature for link prediction between two authors while the current implementation reports Node-sim as the best feature for predicting link between two authors nodes.

**Table 6.** Comparing the performance of link prediction using a single type of feature in the current study to what already was introduced.

| Classifier | Type of feature | Approach | Accuracy% | Precision | Recall | F1-score | AUC |
|---|---|---|---|---|---|---|---|
| RF | RI-sim | Early in [13] | - | 0.002 | 0.014 | 0.007 | 0.49 |
| | | Current approach | | **0.49** | **0.07** | **0.12** | **0.499** |
| | Aff-sim | Early in [13] | - | 0.1 | 0.017 | 0.009 | 0.51 |
| | | Current approach | | **0.50** | **0.97** | **0.66** | **0.5** |
| | I-sum | Early in [13] | - | 0.177 | 0.299 | 0.217 | **0.68** |
| | | Current approach | | **0.52** | **0.59** | **0.55** | 0.52 |
| | Node-sim | Early in [13] | - | 0.02 | 0.005 | 0.003 | 0.5 |
| | | Current approach | | **1.00** | **0.67** | **0.80** | **0.833** |
| KNN | RI-sim | Early in [13] | - | 0.001 | 0.014 | 0.007 | 0.49 |
| | | Current approach | | **0.47** | **0.07** | **0.11** | **0.495** |
| | Aff-sim | Early in [13] | - | 0.003 | 0.014 | 0.007 | 0.5 |
| | | Current approach | | **0.50** | **0.02** | **0.05** | 0.5 |
| | I-sum | Early in [13] | - | 0.172 | **0.295** | **0.214** | **0.67** |
| | | Current approach | | **0.45** | 0.02 | 0.03 | 0.498 |
| | Node-sim | Early in [13] | - | 0.01 | 0.004 | 0.002 | 0.5 |
| | | Current approach | | **1.00** | **0.68** | **0.81** | **0.837** |

## 5. Discussion

Our proposed supervised link prediction based on author node-based features introduced in this paper is a revised and refined version of our early link prediction implementation of this approach introduced in [13]. However, we found that the performance of our link prediction presented in this study outperforms our early implementation of this approach that achieved a modest to poor overall performance. In comparison to the early implementation, our link prediction approach introduced in this study has achieved a high performance link prediction utilizing the same type of features introduced in [13]. This significant improvement in performance might be due to testing our proposed supervised link prediction approach on the two LCC graphs of the two graphs representing ArnetMiner and DBLP co-authorship networks. This makes our proposed approach more realistic for solving the link prediction problem in academic collaboration network in comparison to the early implementation that was only tested on a small subset of the ArnetMiner co-authorship network graph with only 50000 author nodes in non-LCC graph. Thus, it looks like the larger size of co-authorship network with connected nodes leads to better performance of link prediction between authors 'nodes in the collaboration network. Finally, we think that balancing the dataset in this study by sampling a number of negative links equal to the positives ones in the training dataset led to a better performance over our early implementation that suffered from the imbalanced dataset problem [38]. This problem was raised due to using a lower number of positive links in comparison to the number of negative links (the number of positive links was around one third of the number of negative links).

Additionally, we noticed that the performance evaluation results of the early implementation in [13] showed that the sum of research performance indices of two author nodes has the most impact on the performance of supervised link prediction in co-authorship networks. On the other side, the results of our link prediction approach presented in this study show that the three author-based features (RI-sim, AFF-sim, and I-sum) have an equal impact on solving the link prediction problem with less important than the Node-sim feature which has the greatest impact on predicting the potential link between two authors as it was depicted in Tables 3 and 4. We think that the great impact of Node-sim feature on the performance of link prediction approach presented in this study comes from the significant role that the local network topological features extracted by computing the similarity between each two nodes on the SN graph, play on predicting the potential link between two nodes [39].

**6. Conclusions and future work**

In this paper, we have presented a supervised learning approach for predicting potential collaboration in academic co-authorship networks based on author node-based features extracted from such networks. Those features include three types of features. The first type is the features extracted by computing the similarity between the research interests and the similarity between the affiliations of each pair of authors in the co-authorship network. The second type of feature is the features extracted by aggregating the research performance indices of each two authors nodes in the co-authorship network. Finally, the last type of feature is extracted by computing the similarity between each two author nodes in the collaboration network. Our experimental results on the LCC graphs of two graphs representing ArnetMiner and DPLB academic collaboration networks show that our proposed approach achieved outstanding performances. Additionally, the results show almost equal contributions of the similarity between the research interests between two authors, the similarity between their affiliations and the sum of their research performance indices to the link prediction performance, while they show a greater contribution of the similarity score between two authors nodes to the performance of link prediction models/classifiers implemented in this study.

As a future work, we are looking forward to applying our proposed link prediction approach to different academic co-authorship networks and comparing the performance evaluation results. Additionally, we think about the possibility of adopting our approach for performing link prediction on other types of social networks different from the academic ones (e.g., Facebook or X) and check if we can draw the same conclusion. We think that this can be achieved through computing the similarity between posts on Facebook or tweets on X SNs and replacing the research performance indices with different metrics that represent how much the node is influential on such types of networks [40].

**Data and code availability**

The data used in the manuscript is publicly available at no cost [14, 16]. The code of the proposed approach in this study is available at: https://github.com/hsdoaa/co-author_link_prediction_V2

**Disclosure statement**

The authors report there are no competing interests to declare.

**ORCID**

Doaa Hassan https://orcid.org/0000-0002-7358-0830
Mohammad Hasan https://orcid.org/0000-0002-8279-1023